# Combining the target trial and estimand frameworks to define the causal estimand: an application using real-world data to contextualize a single-arm trial

**Authors:** Lisa V Hampson[1], Jufen Chu[2], Aiesha Zia[1], Jie Zhang[2], Wei-Chun Hsu[3], Craig Parzynski[3], Yanni Hao[2], Evgeny Degtyarev[1]

**Affiliations**

1. Novartis Pharma AG, Basel, Switzerland
2. Novartis Pharmaceuticals Corporation, East Hanover, NJ, USA
3. Genesis Research, Hoboken, NJ

**Abstract:** Single-arm trials (SATs) may be used to support regulatory submissions in settings where there is a high unmet medical need and highly promising early efficacy data undermine the equipoise needed for randomization. In this context, patient-level real-world data (RWD) may be used to create an external control arm (ECA) to contextualize the SAT results. However, naive comparisons of the SAT with its ECA will yield biased estimates of causal effects if groups are imbalanced with regards to (un)measured prognostic factors. Several methods are available to adjust for measured confounding, but the interpretation of such analyses is challenging unless the causal question of interest is clearly defined, and the estimator is aligned with the estimand. Additional complications arise when patients in the ECA are eligible for the SAT at multiple timepoints. In this paper, we use a case-study of a pivotal SAT of a novel CAR-T therapy for heavily pre-treated patients with follicular lymphoma to illustrate how a combination of the target trial and the ICH E9(R1) estimand frameworks can be used to define the target estimand and avoid common methodological pitfalls related to the design of the ECA and comparisons with the SAT. We also propose an approach to address the challenge of how to define an appropriate time zero for external controls who meet the SAT inclusion/exclusion criteria at several timepoints. Use of the target trial and estimand frameworks facilitates discussions amongst internal and external stakeholders, as well as an early assessment of the adequacy of the available RWD.

1. Introduction

Randomized controlled trials (RCTs) are considered to be the gold standard for evaluating the efficacy and safety of new treatments and for supporting informed decisions by regulators, payers, physicians and patients. Randomization ensures that in large samples, the two groups of patients will be balanced with respect to measured and unmeasured prognostic factors and, hence, with respect to their risks of any type of health outcome. Consequently, a comparison between treatment groups provides an unbiased estimate of the causal treatment effect[1].

Single-arm trials (SATs) are often performed in oncology to support health-authority approval. A retrospective study of European Marketing Authorizations between 2010 and 2019, identified 22 initial marketing authorizations based on SATs[2]. A review of US Food and Drug Administration (FDA) oncology and hematology approvals between 2008 and 2016 identified 25% of approvals as being primarily supported by SATs[3].

Real-world data (RWD) are data relating to patients' health status and/or the delivery of health care which are routinely collected when patients interact with the healthcare system. Therefore, by definition RWD can be obtained from a variety of sources such as electronic health records (EHRs), insurer and/or provider administrative claims data, disease or patient registries, or chart reviews. Real-world evidence (RWE) describes any analysis using RWD[4]. There has been a long history of using RWD for post-approval safety and effectiveness studies of medical products. More recently, there is growing interest in the use of RWE to support pre-authorization regulatory decisions. Much momentum comes from regulators' interest in leveraging RWE for decision-making as evidenced by the 21st Century Cure Act in the US, and in Europe the HMA – EMA Joint Big Data Task Force and the European Medicine Agency (EMA) regulatory science 2025 strategic reflection. Existing use cases include leveraging RWD for regulatory submission of supplemental indications and to support full registration after an initial accelerated or conditional approval of a medical product. The use of patient-level RWD to create an external control arm (ECA) represents an important application as comparisons between a SAT and ECA can be useful to contextualize SAT results. In recent years, the availability of RWD (e.g., EHRs with linked clinico-genomics data) for highly targeted populations with molecular subsets have also facilitated the creation of ECAs.

Naive comparisons between the treated cohort of a SAT and an ECA can yield biased estimates of the causal treatment effect due to systematic differences between groups with respect to measured and unmeasured prognostic factors. Different methodologies have been proposed to adjust for measured confounding and used to support regulatory submissions for new treatments in US and EU[5,6,7]. However, the interpretation of such analyses may be challenging if the causal question of interest is not clearly defined, or the estimator is not aligned with the estimand. Hernán and Robins[8] highlighted that causal analyses of observational data need to be evaluated with respect to how well they emulate a particular target RCT, i.e., a hypothetical RCT that would answer the question of interest if conducted. They outlined a framework to make the target trial explicit by providing a structured approach to define the question of interest and to avoid common methodological pitfalls.

In this paper, we discuss the application of the target trial framework when comparing a SAT with an ECA to support regulatory submissions. Section 2 introduces the motivating example of a SAT of tisagenlecleucel in follicular lymphoma. In Section 3, we describe how the indirect comparison of the SAT and ECA was planned using a combination of the target trial and ICH E9(R1) estimand frameworks, while Section 4 elaborates on how we addressed the challenge of defining an appropriate time zero for external controls who meet the SAT eligibility criteria at several timepoints, and how we used the selected data to estimate the causal estimand. We present the results of our case-study in Section 5 before concluding with a discussion of our practical experiences of applying the combined target trial and estimand frameworks and their potential value for drug development in Section 6.

## 2. Motivating example
### 2.1. The ELARA trial: a single-arm pivotal Phase II study

Follicular lymphoma is a non-Hodgkin lymphoma that is generally considered indolent, but the disease remains incurable and the majority of patients eventually relapse. Patients with relapsed or refractory (r/r) follicular lymphoma will experience progressively shorter progression-free survival (PFS) to subsequent treatments, which has been shown to decrease from 6.6 years after the first line of therapy (LoT) to 1.5 years and 10 months after the second and third lines of therapy, respectively[9]. Hence, r/r follicular lymphoma remains the leading cause of mortality for patients with this rare disease and represents an unmet medical need[10].

The ELARA trial is an ongoing, single-arm, global, multicenter, Phase II trial to determine the efficacy and safety of tisagenlecleucel (a novel cell therapy) in adult patients with r/r FL who were r/r to second or later line therapy[11,12]. A total of 98 patients were enrolled, of whom 97 were infused with tisagenlecleucel. A study schema is presented in Figure 1.

[Insert Figure 1 here]

During the national regulatory authority protocol review process, an EMA rapporteur highlighted the importance of having an ECA with patient-level data to support the review of any potential submission based on the ELARA trial. Consequently, RWE was derived using two external data sources to provide stakeholders with evidence on the magnitude of the effect of tisagenlecleucel efficacy compared with the standard of care (SoC).

### 2.2. Datasets used to construct two external control arms for ELARA

A non-interventional retrospective chart review study, titled *A Retrospective Cohort Study of Treatment Outcomes Among Adult Patients with Refractory or Relapsed Follicular Lymphoma (ReCORD-FL)* (hereafter referred to as "ReCORD"), was conducted to provide comparative, contextual evidence to the existing data on the efficacy of tisagenlecleucel from ELARA. Patient-level data were collected from centers in Europe and North America[13]. To obtain an adequate sample size and to include patients treated with different regimens reflecting evolving management strategies, data were collected from patients with r/r follicular lymphoma treated between 1998 and 2020. No initial diagnoses before January 1, 1998 were permitted as a key treatment for r/r follicular lymphoma, rituximab, was approved by the FDA and EMA in 1997 and 1998, respectively.

Wherever feasible, the ReCORD study adopted the same eligibility criteria as ELARA. As of cutoff date of Dec 31, 2020, 187 patients fitting the study eligibility criteria were enrolled. For each eligible patient and each line of therapy (LoT) from first diagnosis onward, the physician recorded the patient's best clinical response to the treatment line, dates of progression, death, or start of new anti-cancer therapy.

A second non-interventional retrospective study utilizing the Flatiron Health Research Database (FHRD), was conducted to contextualize the ELARA trial. The FHRD is a nationwide US EHR-derived de-identified database originating from approximately 280 cancer clinics (~800 sites of care)[14]. It is composed of de-identified patient-level structured and unstructured data, curated via technology-enabled abstraction[15, 16]. The data delivery included patients diagnosed with follicular lymphoma between January 1, 2011 and June 30, 2020 who had received at least 3 lines of therapy. The majority of patients in the database originate from community oncology settings.

Wherever feasible, the ELARA eligibility criteria were applied to the FHRD. As of cutoff date of June 30, 2020, 98 patients satisfying the ELARA eligibility criteria and receiving treatment between January 1, 2014 and April 1, 2020 were selected. For each LoT, progression and response were abstracted based on clinician assessments as documented in the EHR. Treatment and mortality information were collected through an amalgamation of structured data elements, unstructured documents, and linking to external mortality sources and the Social Security Death Index where necessary. For both the ReCORD and FHRD studies, key baseline and clinical variables were collected or derived at the start of each LoT.

### 3. Defining the target estimand for studies with a RWD-based ECA

#### 3.1 *Motivation for combining the target trial and ICH E9(R1) estimand frameworks*

The process of designing an ECA for a SAT begins with specifying the causal question that we want to answer with the comparison of these two (non-randomized) groups. Shaping a relevant and feasible question requires input from several stakeholders, including clinicians and statisticians, and there are several frameworks which can structure this process.

The PICO aid[17] plays an important role in evidence-based medicine, often informing the search criteria for quantitative systematic reviews and how selected studies are combined[18]. The PICO acronym itemizes four key elements of a clinical question about treatments: patient **P**opulation, **I**ntervention, **C**ontrol, and **O**utcome(s). Extensions such as PICOT and PICOS go further to make explicit the role of **T**ime (i.e., length of follow-up) and **S**tudy design, respectively, in a research question. Clearly, the PICO elements overlap with the attributes of an estimand as defined by the ICH E9(R1) estimand framework[19]. However, we prefer the estimand framework since this goes even further to capture how the effect of a treatment (vs control) will be summarized, and outlines strategies for handling intercurrent events which either affect the presence or interpretation of relevant data. The estimand framework is relevant for RCTs as well as SATs[19] and clinical trials run through registries[20].

Beyond clinical trials, the ENCePP Guide on Methodological Standards in Pharmacoepidemiology advises on how to set the research question for non-interventional studies[21]. A number of frameworks and tools have also been proposed for studies generating RWE[22,23]. Section 1 of this paper highlighted the target trial framework[8] as a tool that can be used to ensure the design and analysis of an observational study is aligned with a well-posed causal question. This framework guides the researcher to carefully define key elements of the protocol of the 'target' RCT that would answer the causal question of interest. Several of these elements specify ICH E9(R1) estimand attributes such as patient population, treatment strategies, outcome variable and summary measure. Additional protocol sections such as 'follow-up period' facilitate careful consideration of how to define start of follow-up (also referred to as 'time zero') for patients; while this may be straightforward for RCTs, since start of follow-up usually coincides with time of randomization, defining time zero is often more challenging for non-interventional studies making secondary use of RWD. However, the target RCT protocol defined by Hernán and Robins[8] does not include a section explicitly detailing how intercurrent events would be handled. Several authors have proposed combining the target trial and ICH E9(R1) estimand frameworks[24, 25] for clinical studies using RWD. In particular, Hampson et al.[24] propose extending the protocol of the target RCT protocol to include a section on handling intercurrent events, in order to facilitate a more granular approach to handling intercurrent events and communication of these strategies. Since many clinical trial teams are already familiar with the estimand framework through its application to RCTs, we decided to adopt this combination of the target trial and estimand frameworks for the ELARA trial. We describe this approach in more detail below.

### 3.2 *Applying the target trial and estimand frameworks to the ELARA trial*

Table 1 illustrates how the combined target trial and estimand frameworks were applied to design a real-world- (RW) based ECA for the ELARA trial. Broadly speaking, for each key efficacy endpoint, the question of interest selected by the team was the marginal causal effect of prescribing tisagenlecleucel vs SoC in the patient population who participated in the ELARA trial; in common terminology, this is an 'average treatment effect on the treated' (ATT). The left-hand column of Table 1 defines the protocol of the RCT that we would perform to answer this question (referred to as the 'target RCT'). Key points to note are that since we are targeting an ATT, the eligibility criteria of the target

RCT are of those of ELARA. We aim to emulate an active controlled randomized trial comparing the tisagenlecleucel treatment strategy evaluated in ELARA (and illustrated in Figure 1) with the SoC per physician's choice. In this indication, the number of prior lines of therapy at baseline is a well-understood strong prognostic variable. In terms of the key efficacy endpoints, the most relevant intercurrent event in the target RCT would be initiation of a new anticancer therapy. For complete response (CR), if a patient hasn't achieved response prior to starting a new therapy they would be regarded as a non-responder. For PFS, initiation of a new therapy would be handled through a hypothetical strategy where a patient would be censored at the time of initiating a new anticancer therapy if no progression or death is observed before the new anticancer therapy. Meanwhile, for overall survival (OS), it is handled using a treatment policy approach, recording the time to all-cause mortality regardless of whether a patient deviates from their randomized strategy. The causal contrasts are difference in probabilities (CR) and hazard ratios (PFS and OS).

[Insert Table 1 here]

The right-hand columns of Table 1 summarize the RCT that we *can* emulate given the data available from ELARA and the external controls. Comparing this with our target RCT highlights limitations of the RWD. In particular, while we can emulate the target population based on key ELARA eligibility criteria as envisioned in future drug label criteria, not all criteria were captured in the RWD (e.g. ECOG or lab results were not recorded or incomplete). However, based on clinical assessment it was concluded that the impact of these criteria on the adequacy of the ECA is likely to be limited and does not prevent meaningful comparisons. Another potential limitation is that CR and progression in the RWD were assessed per real-world regular practice and not following the Lugano classification 2014 criteria[26] as used in ELARA. Regarding this, a subgroup analysis for patients treated after 01-Jan-2014, to coincide with the introduction of the new Lugano response criteria, was conducted. Moreover, this limitation does not impact OS. A third limitation is that while the date of treatment initiation is available for all patients, the date of prescription is unknown for RW patients. However, in the context of currently available therapies and a serious condition such as follicular lymphoma, the lag between prescription and start of treatment is likely to be short, meaning the date of treatment initiation will be an adequate approximation for the date of prescription. Since the date of initiation of all treatments is recorded for RW patients, the key intercurrent event of initiation of a new anticancer therapy is captured. We can apply the same strategy as would be used in the target RCT to handle this event when comparing the ELARA and external controls with regards to CR and OS. However, progression dates were unavailable for many patients in ReCORD, so the comparison of PFS was performed by considering a new anticancer therapy as an event for patients in both ReCORD and ELARA. If no progression, death or starting a new anticancer therapy occurred, patients are censored at the last contact date. For FHRD, PFS was handled in the same way as the target RCT. If patients have not had disease progression before the start of new anticancer therapy, they were censored at the date of the starting new anticancer therapy, or last clinic note date if the selected LoT was the last LoT.

Section 4 describes how we attempted to emulate the treatment assignment and follow-up strategies of the target RCT using a propensity-score based approach to select one LoT per RW patient and weight outcomes on the selected line to mimic randomization.

## 4. Selecting one LoT per RW patient and estimating the estimand

As mentioned in Section 2, the ELARA trial was open to patients who were r/r to a second or later line of systemic therapy, with no upper limit placed on the number of previous lines of therapy. We refer to a patient's 'index date' as the calendar time of their first qualifying LoT at which they satisfy all eligibility criteria. The median number of previous LoT on entry to ELARA was 4 (range: 2 – 13). In the ReCORD cohort, 80% of patients had received 2 previous LoT at the time of first meeting all eligibility criteria (range: 2 – 5 lines), while for the FHRD cohort, 100% of patients first met the eligibility criteria after 2 previous LoT (range: 3 – 5 lines).

Another key difference between the ELARA and RW patients is that CR and PFS were assessed in the trial only for a single LoT, whereas longitudinal data were available on these endpoints for RW patients who were followed and remained eligible across multiple LoT. These differences in data structure are common for SATs in late-stage r/r disease with RW-based ECAs.

If observational data from the same source were to be used to emulate both arms of a target RCT and patients would meet the eligibility criteria at multiple LoT, valid approaches to LoT selection would include selecting a single eligible line or using data from all eligible LoT to emulate multiple nested RCTs[8]. However, our case study differs from that setting because there are systematic differences in data capture between the ELARA trial and the RWD. In the case of a SAT in late-line therapy with RW external controls, Backenroth[27] evaluates several approaches for defining start of follow-up for the external controls with respect to bias for the treatment effect; these approaches include: selecting the last eligible LoT; selecting a line at random from a patient's eligible LoT; and using information on all eligible LoT. In the ELARA example, for ease of analysis and interpretation, it was decided to mimic the structure of the ELARA dataset by selecting one eligible LoT per RW patient. This is also closely aligned with the structure of a future RCT that would compare responses on tisagenlecleucel and SoC on the first received LoT in the trial. While this approach sacrifices some information, it has the advantage of avoiding correlations between repeated measurements on the same RW patient and ensures individual trial participants and external controls contribute similar amounts of statistical information to the final analysis. The challenge then is how to select one eligible LoT per RW patient.

We took a novel approach to select a single LoT. First the ELARA data and the available RWD (comprising observations on all eligible LoT for all eligible patients) are used to fit a model for patients' probability of being in the ELARA trial; we refer to this as the propensity score model. One could model these data using a mixed effects logistic regression, with baseline covariates represented by fixed effects in the linear predictor and a random subject effect included for each RW patient to capture the correlation between their longitudinal data. These random effects could be assumed to follow a normal distribution centered at zero, or a distribution with heavier tails (such as a t-distribution on few degrees of freedom) if outliers are considered possible. Since each SAT participant is followed-up through only one LoT, their subject effect is set to zero. While such a mixed-effects model would precisely capture the structure of the ELARA data and external controls, one would need to write bespoke code to fit it and there is the potential for convergence issues if few RW patients have data available on multiple LoT. With these challenges in mind, for simplicity we modelled the probability of the $i$th patient being in the ELARA trial at their $j$th eligible LoT using the fixed-effects logistic regression shown below:

$$\log\left(\frac{e_{ij}}{1 - e_{ij}}\right) = \mu + \sum_{k=1}^{K} \beta_{ijk} X_{ijk}$$

where $X_{ijk}$ is the value of the $k$th covariate for the $i$th patient measured at the time of initiating their $j$th eligible LoT. Ten covariates were pre-specified by clinical experts before seeing the outcome data for RW patients including age, region (race was used in FRHD since region was unavailable), gender, history of autologous HSCT, number of prior lines of therapy, disease stage at initial diagnosis, time from initial diagnosis to index treatment, sites of nodal involvement, double refractory status and POD24. The logistic regression was fitted using a generalized estimating equations (GEE) approach, using a robust sandwich variance estimator to ensure the standard errors of model parameters reflect the correlation between longitudinal measurements on the same patient. However, in what follows, only the point estimates of propensity scores are used to inform LoT selection, which ignore the correlation between repeated measurements on RW patients. Fitting the propensity score model generates propensity score estimates $\{\hat{e}_{ij}; i = 1, \cdots, N_{RW}, j = 1, \cdots L_i\}$ for the RW cohort, where N$_{RW}$ denotes the number of RW patients eligible for at least one LoT, meaning L$_i$ ≥ 1 for each $i = 1, \cdots, N_{RW}$. For each RW patient, we then selected the LoT associated with their highest propensity score. Intuitively, we interpret this as the line at which the RW patient is most similar (in terms of their baseline covariates, including number of previous lines of therapy) to the ELARA cohort. We note that for patients eligible at multiple LoT, the selection process is not directly influenced by a patient's number of eligible lines; instead, selection is more closely linked with our objective of trying to minimize confounding between groups. In addition, our proposed approach preserves the sample size of the RW cohort.

The dataset formed by the ELARA cohort and the RW cohort with one selected LoT per patient is the final analysis dataset. Note that imbalances between groups will likely still persist in this dataset. To resolve these, we use the selected dataset to re-fit propensity score model (1) using least squares estimation, since all patients now contribute only one observation to the analysis. The propensity score estimates can then be used in a PS-based analysis to adjust for confounding. In Section 3, we defined the target estimand as the average effect of prescribing tisagenlecleucel as compared to SoC in those who participated in the ELARA trial. To estimate this, we reweight patients to create tisagenlecleucel and SoC groups with covariate distributions identical to the distribution seen in the ELARA trial. Clearly, we don't need to reweight the ELARA cohort to create a tisagenlecleucel group fulfilling this objective, so all patients in this group can be assigned a weight of 1. Meanwhile, assigning the $i$th RW patient a weight equal to $\hat{e}_i/(1 - \hat{e}_i)$ creates a group with the same covariate distribution as the ELARA cohort.

The weighted datasets can be used to compute estimates of causal effects by comparing weighted aggregate statistics (e.g. the difference in the weighted proportion of responders for binary endpoints). Alternatively, we can extract treatment effect estimates from (semi-)parametric models fitted allocating patients different weights, e.g. estimating the log-HR by fitting a weighted Cox regression with treatment as the only regressor and treating the weights $\hat{e}_i/(1 - \hat{e}_i)$ for $i = 1, \cdots, N_{RW}$, as replication weights. The standard error of the causal treatment effect estimator should account for the fact that patient weights are estimated from the data. Austin[28] compared three approaches (naive-model based variance estimator, robust sandwich-type variance estimator, and bootstrap) for estimating the variance of the treatment effect estimate obtained from an inverse probability treatment weighting analysis. The author found that the bootstrap estimator provides the most robust estimate of the standard error of the hazard ratio from a Cox proportional-hazards model. In the ELARA case-study, we used non-parametric bootstrap to estimate the standard error of treatment effect estimates based on 10,000 bootstrap samples[29] randomly drawn (with replacement) from the dataset obtained after selecting one LoT per RW patient.

## 5. Application to the ELARA trial

In this section, we provide the application of the methods described in the previous section to the ELARA trial using ReCORD and Flatiron data as ECAs. Since our primary aim was to use these two ECAs to contextualize the ELARA trial and understand the magnitude of the effect of tisagenlecleucel versus SoC, we summarize the results of comparative analyses with point estimates and confidence intervals rather than p-values.

### 5.1 *Results using ReCORD data as the ECA*

Included in the analysis were a total of 97 patients (out of 98 enrolled) from ELARA and 143 patients (out of 187 recruited) from the ReCORD study who had complete data on key baseline covariates; this sample consisted of 326 LoT for 143 ReCORD patients. As described in Section 4, to mimic the structure of the ELARA dataset, a propensity score-based approach was used to select one LoT per RW patient.

After line selection, we observed that the median number of prior lines of therapy in ReCORD is close to ELARA (3 vs 4), with 23.1% of ReCORD patients having received more than 4 prior lines of therapy (vs 28.9% in ELARA). However, imbalances between groups still persisted in many baseline covariates. Thus, as described in section 4, the weighting by odds approach was used to adjust for potential confounding. Table 2 shows the comparison of baseline variables between ReCORD patients (at their selected LoT) and ELARA patients before and after ReCORD patients are weighted by their odds of being in ELARA. Standardized mean differences (SMD) between the two cohorts were assessed for pre-weighted and post-weighted data. The weighting by odds analysis appears to be reasonable to adjust for imbalances of measured baseline covariates between groups as suggested by absolute SMDs less than 25%[30]. The estimated propensity scores at the selected LoT for ReCORD patients and ELARA patients are shown in the upper panels of Figure 2, with a smoothed estimate of the density overlaid by exposure groups. The lower panel of Figure 2 repeats this comparison but this time weighting ReCORD patients by their odds of being enrolled in ELARA. The weighting by odds method appears to have adequately balanced the propensity score distribution between these two groups.

[Insert Table 2 here]

[Insert Figure 2 here]

Table 3 shows the estimates of the causal effect of prescribing tisagenlecleucel versus SoC before and after weighting by odds. Figure 3 shows the Kaplan-Meier curves for OS and PFS after weighting. The median follow-up time (defined as time to death or last follow-up date) was 15 months for ELARA, and 22 months in the weighted sample for ReCORD (at the selected LoT). For comparative analyses, the Kaplan-Meier and Cox regression results are based on survival data within the first 24 months, and patients with survival data beyond 24 months were censored at 24 months. A clinically meaningful and consistent improvement for all endpoints was observed before and after weighting in ELARA vs ReCORD. The results from subgroup analysis for patients treated after 01-Jan-2014 are generally consistent with results in the main analysis presented above.

[Insert Table 3 here]

[Insert Figure 3 here]

### 5.2 *Results using FHRD data as the ECA*

A total of 97 patients from ELARA and 98 patients representing 149 LoTs from FHRD were included in the analysis. One patient from ELARA with missing stage at initial diagnosis was excluded to achieve model convergence. Similar to ReCORD, one LoT was selected per patient. After LoT selection, the median number of prior lines of therapy in FHRD was smaller than ELARA (2 vs 4), with 6.1% of FHRD patients having received more than 4 prior lines of therapy (vs 28.9% in ELARA). Post selection of index line in FHRD, weighting by odds was implemented to account for residual confounding using the same set of prognostic variables except history of autologous HSCT and sites of nodal involvement at initial diagnosis. These two variables were extremely imbalanced and were excluded to achieve better balance for all other variables. Given ELARA patients had worse prognosis with respect to these variables, not adjusting for them represents a more conservative scenario where the endpoints favor the control group. Weighting by odds appears to adequately reduce imbalances of measured baseline covariates between groups since the majority of the SMDs are less than 25% and distributions of the propensity score are similar after weighting (as shown in Figure 4).

[Insert Figure 4]

Table 4 shows the estimates of the causal effect of prescribing tisagenlecleucel versus SoC before and after weighting by odds. The median follow-up time (defined as time to death or last activity date) was 15 months for ELARA, and 14 months in the weighted sample for FHRD (at the selected LoT). There was a consistent trend towards greater CRR, ORR, OS and PFS in favor of tisagenlecleucel versus SoC before and after weighting in ELARA vs FHRD.

[Insert Table 4]

## 6. Discussion

In the ELARA example, members of the clinical trial team found that a combination of the target trial and estimand frameworks was helpful to formulate a well-posed and feasible causal question and to align the estimator with the estimand. Table 1, which compares the protocol of the RCT we want to emulate with that of the one we can emulate, was used in communications with clinical trial team members and health authorities, and facilitated transparent discussion of the feasibility and value of the RW-based ECAs. To our knowledge, this is the first time the target trial framework has been used in a regulatory submission with RWE.

Several learnings arose from this case study. Firstly, if planning an ECA, it is best to use the target trial and estimand frameworks to define the estimand as early as possible, preferably before drafting the protocol for the indirect comparison and RWD collection form. This will also allow a meaningful discussion with health authorities about the key questions of interest and the opportunity to address these questions with RWE. Secondly, use of target trial and estimand frameworks can facilitate internal alignment on study objectives and an early assessment of whether key question(s) of interest can be reliably addressed with the RWD source. In the ELARA example, the combined frameworks helped to clarify that the key question was efficacy from the time of enrollment rather than from tisagenlecleucel infusion. Moreover, the frameworks also facilitated discussions about other important aspects such as line selection for RW patients. An additional learning concerned the challenges associated with selecting a single LoT for RW external controls and highlighted the approach proposed in Section 4 as a feasible solution.

Our final key learning is that it is important to align ahead of time with clinical collaborators and health authorities on which baseline covariates are anticipated to be strong prognostic and/or predictive factors, to ensure all relevant patient information is extracted. However, while it is important to follow the same principles for RWD analyses as for clinical trials and prespecify the statistical analysis plan, some flexibility may be required from all stakeholders. For example, while 'prior autologous haematopoietic stem cell transplantation' was pre-specified for inclusion in the propensity score model, analyses revealed that 2% of patients in the FHRD cohort had received prior transplantation compared with 37% of patients in ELARA. Therefore, use of this covariate in the model was not considered meaningful. Since prior transplantation is associated with worse outcomes, the analysis without adjustment for this variable was still performed as the results could be interpreted as being conservative from the tisagenlecleucel perspective. Additionally, post-hoc sensitivity analyses were performed focusing on the subgroup of patients without prior transplantation.

Two analytical challenges that we haven't discussed in detail are: a) how to handle missing data on key baseline covariates or efficacy outcomes when formulating a RW-based ECA; and b) how to evaluate the robustness of conclusions to unverifiable assumptions. For ELARA, we took a rather simple approach of handling missing data via a complete-case analysis but alternative approaches, such as multiple imputation[31], could have been considered. In terms of assessing the robustness of results, having a precisely defined estimand ensures sensitivity analyses can be aligned with the question of interest. In the field of epidemiology, a large body of literature exists on quantitative bias assessment and metrics quantifying the sensitivity of conclusions to unmeasured confounding exist for binary and time-to-event endpoints[32, 33]. Future work would develop best practices for sensitivity analyses for ECAs and extend methods to accommodate the range of endpoints we might encounter in clinical trials.

In conclusion, we think combining the target trial and estimand frameworks can be useful when planning a SAT with an ECA as it offers a common language to discuss existing complexities with all relevant stakeholders including health authorities and payers. Experience with these frameworks in clinical studies with RWD is still limited, and their use in the oncology setting is being explored in dedicated industry working groups. More published case-studies will be useful for guiding researchers.

**Acknowledgements**

We are grateful to acknowledge helpful discussions with Hemanth Kanakamedala which informed the development of methods described in this manuscript.

**References**

1. Collins R, Bowman L, Landray M, Peto R. The Magic of Randomization versus the Myth of Real-World Evidence. *New England Journal of Medicine* 2020; 382: 674-678
2. Tenhunen O, Lasch F, Schiel A, Turpeinen M. Single-Arm Clinical Trials as Pivotal Evidence for Cancer Drug Approval: A Retrospective Cohort Study of Centralized European Marketing Authorizations Between 2010 and 2019. Clin Pharmacology & Therapeutics 2020;108(3):653-660.
3. Zhou J, Vallejo J, Kluetz P, Pazdur R, Kim T, Keegan P, Farrell A, Beaver JA, Sridhara R. Overview of oncology and hematology drug approvals at US Food and Drug Administration between 2008 and 2016. *Journal of National Cancer Institute* 2019;111, djy130.
4. Food and Drug Administration. Real-World Evidence. Available at: https://www.fda.gov/science-research/science-and-research-special-topics/real-world-evidence
5. FDA Center for Drug Evaluation and Research.Multi-discipline review. NDA 212306 XPOVIO (Selinexor). Available at: https://www.accessdata.fda.gov/drugsatfda_docs/nda/2019/212306Orig1s000MultidisciplineR.pdf ; 2019a
6. FDA Center for Drug Evaluation and Research. Multi-discipline review. NDA 212018 BALVERSA (Erdafitinib). Available at: https://www.accessdata.fda.gov/drugsatfda_docs/nda/2019/212018Orig1s000MultidisciplineR.pdf; 2019b
7. FDA Center for Drug Evaluation and Research. Multi-discipline review. BLA 761163 Monjuvi (Tafasitamab-cxix). Available at: https://www.accessdata.fda.gov/drugsatfda_docs/nda/2020/761163Orig1s000MultidisciplineR.pdf; 2019c
8. Hernán MA, Robins JM. Using Big Data to Emulate a Target Trial When a Randomized Trial Is Not Available. *American Journal of Epidemiology* 2016;183(8):758-64.
9. Link BK, Day B-M, Zhou X, Zelenetz AD, Dawson KL, Cerhan JR, Flowers CR, Friedberg JW. et al. Second-line and subsequent therapy and outcomes for follicular lymphoma in the United States: data from the observational National LymphoCare Study. *British Joural of Haematology* 2019; 184:660-663. doi:10.1111/bjh.15149.
10. Sarkozy C, Maurer MJ, Link BK, Ghesquieres H, Nicolas E, Thompson CA, Traverse-Glehen A, Feldman AL, Allmer C, Slager SL, Ansell SM, Habermann TM, Bachy E, Cerhan JR, Salles G.. Cause of death in follicular lymphoma in the first decade of the rituximab era: a pooled analysis of French and US cohorts. *Journal of Clinical Oncology* 2019;37:144-152.
11. Fowler NH, Dickinson M, Dreyling M, Martinez-Lopez J, Kolstad A, Butler JP, Ghosh M, Popplewell LL, Chavez JC, Bachy E et al. Efficacy and Safety of Tisagenlecleucel in Adult Patients with Relapsed/Refractory Follicular Lymphoma: Interim Analysis of the Phase 2 Elara Trial. *Blood* 2020; 136 (Supplement 1): 1–3. doi: https://doi.org/10.1182/blood-2020-138983
12. Schuster S, Bishop MR, Tam CS, et al: Tisagenlecleucel in adult relapsed or refractory diffuse large B-cell lymphoma. *New England Journal of Medicine* 2019; 380:45-56
13

**Table 1:** Application of the target trial and estimand frameworks to design an ECA based on the ReCORD study and the comparison of this with the ELARA trial cohort. The same considerations applied when constructing an ECA from the FHRD.

| Component | Target RCT to be emulated | RCT that can be emulated using ELARA and external RWD | |
|---|---|---|---|
| | | **ELARA** | **RW cohort** |
| Objective | To evaluate the efficacy of tisagenlecleucel as compared to the current SoC as measured by complete response (CR), overall survival (OS), and progression free survival (PFS) | | |
| Patient Population | Eligibility criteria of ELARA | Same as target RCT | Eligibility criteria of ELARA that are feasible to implement in a retrospective assessment of the RWD |
| Treatment | Optional bridging therapy and lymphodepleting therapy followed by tisagenlecleucel infusion vs SoC | | |
| Treatment assignment | Block randomized to either tisagenlecleucel arm or SoC arm. | Tisagenlecleucel infusion after optional bridging therapy and Lymphodepleting therapy | One eligible LoT per patient in the external source was systematically selected based on the highest propensity score to be in ELARA. See Section 4 for details. |
| Variables | CR and progression are assessed per Lugano classification 2014[26] until patient initiates a new anticancer treatment. CR is best overall response. If no response is reported before progression or time of new therapy, patient is regarded as non-responder. PFS is time to the first documented progression or death from any cause. Initiation of a new anticancer therapy would be handled through a hypothetical strategy: if a patient | CR and OS as target RCT. For comparison with ReCORD, PFS is performed by considering new anticancer therapy as an event. For comparison with FHRD, PFS is the same as target RCT. | CR and progression were based on real-world response criteria. For FRHD, only abstracted response with non-missing value were considered, and only patients with at least >=1 non-missing value for abstracted response were included for CR analysis. For ReCORD, a subgroup analysis for patients treated after 01-Jan-2014 was conducted as year of introduction of Lugano response criteria. In ReCORD, progression dates were unavailable for many patients, so consider new anticancer therapy as an event for PFS. In FHRD, PFS was handled using the same hypothetical strategy as target RCT. OS same as in target RCT. |

| Component | Target RCT to be emulated | RCT that can be emulated using ELARA and external RWD | |
|---|---|---|---|
| | | ELARA | RW cohort |
| | does not have progression or death prior to initiation, PFS would be censored at the time of initiation. OS is time to death from any cause regardless of treatment | | |
| Start of follow-up | Date of randomization, regarded as date of prescription | Date of enrolment into ELARA, regarded as date of prescription | Start date of SoC |
| Intercurrent event(s) | Initiation of new anticancer therapy before progression. CR: ICE reflected in Variable PFS: Hypothetical strategy (reflected in Variable) OS: Treatment policy strategy (reflected in Variable) | Same as target RCT for CR and OS PFS comparison with ReCORD: Composite strategy PFS comparison with FHRD: the same as target RCT. | |
| Performed comparison | Relative benefit after prescribing tisagenlecleucel vs SoC | Relative benefit after prescribing tisagenlecleucel vs after being treated with SoC assessed by weighting by odds, as described in Section 4. | |
| Causal effect | Effect of prescribing tisagenlecleucel vs SoC in patients meeting ELARA inclusion/exclusion criteria | The effect of prescribing tisagenlecleucel vs SoC is assessed in patients who participated in ELARA | |
| Causal contrast (i.e. Summary measure) | Binary endpoints: Difference in marginal response probabilities on | Same as target RCT | |

| Component | Target RCT to be emulated | RCT that can be emulated using ELARA and external RWD ||
| | | ELARA | RW cohort |
|---|---|---|---|
| | tisagenlecleucel vs SoC<br><br>Time-to-event (TTE) endpoints: Marginal HR | | |
| Analysis | Binary endpoints: Difference in response rate between arms<br><br>TTE endpoints: HR obtained from Cox regression | Binary endpoints: Difference in weighted proportion of responders on tisagenlecleucel vs SoC.<br><br>TTE endpoints: HR obtained from a weighted Cox regression model<br><br>For both types of analysis, weights are calculated using the methodology described in Section 4. 95% CIs are calculated using bootstrap resampling. ||

**Table 2.** Baseline variables for ReCORD and ELARA before and after weighting at selected LoT

| Baseline variable Statistics | ELARA N=97 | Before weighting ReCORD N=143 | \|SMD\| | After weighting ReCORD N=99[a] | \|SMD\| |
|---|---|---|---|---|---|
| **Included in Propensity Score Model** | | | | | |
| **Age at treatment initiation (years)** | | | | | |
| n | 97 | 143 | 0.326 | 99 | 0.038 |
| Mean(SD) | 56.5 (10.40) | 60.1 (11.72) | | 56.1 (11.52) | |
| Median | 58 | 60 | | 56.3 | |
| Min-Max | 29-73 | 25-86 | | 25-86 | |
| **Age at treatment initiation category - n (%)** | | | | | |
| <65 years | 73 (75.3) | 89 (62.2) | 0.284 | 76 (76.7) | 0.034 |
| ≥65 years | 24 (24.7) | 54 (37.8) | 0.284 | 23 (23.3) | 0.034 |
| **Gender - n (%)** | | | | | |
| Female | 33 (34.0) | 61 (42.7) | 0.178 | 30.8 (31.1) | 0.063 |
| Male | 64 (66.0) | 82 (57.3) | 0.178 | 68.3 (68.9) | 0.063 |
| **Region - n (%)** | | | | | |
| Europe | 44 (45.4) | 90 (62.9) | 0.358 | 41.4 (41.8) | 0.072 |
| RoW | 53 (54.6) | 53 (37.1) | 0.358 | 57.6 (58.2) | 0.072 |
| **Prior Auto-HSCT - n (%)** | | | | | |
| Yes | 36 (37.1) | 53 (37.1) | 0.001 | 36.1 (36.5) | 0.013 |
| No | 61(62.9) | 90 (62.9) | 0.001 | 62.9 (63.5) | 0.013 |
| **Number of previous lines of systemic treatment** | | | | | |
| n | 97 | 143 | 0.117 | 99 | 0.104 |
| Mean (SD) | 3.9 (1.78) | 3.7 (2.05) | | 4.1 (2.25) | |
| Median | 4 | 3 | | 4 | |
| Min-Max | 2-13 | 2-10 | | 2-10 | |
| **Number of previous lines of systemic treatment - n (%)** | | | | | |
| 2-4 | 69 (71.1) | 110 (76.9) | 0.132 | 69.9 (70.6) | 0.011 |
| >4 | 28 (28.9) | 33 (23.1) | 0.132 | 29.1 (29.4) | 0.011 |
| Disease stage at initial FL diagnosis – n(%) | | | | | |
|   Stage I | 6 (6.2) | 10 (7.0) | 0.033 | 4.7 (4.7) | 0.064 |
|   Stage II | 13 (13.4) | 13 (9.1) | 0.137 | 9.5 (9.6) | 0.12 |
|   Stage III | 21 (21.6) | 26 (18.2) | 0.087 | 25.4 (25.7) | 0.095 |
|   Stage IV | 57 (58.8) | 94 (65.7) | 0.144 | 59.4 (60.0) | 0.026 |
| Months between initial FL diagnosis and initiation of treatment | | | | | |
|   n | 97 | 143 | 0.099 | 99 | 0.005 |
|   Mean (SD) | 77.3 (56.33) | 72.1 (48.53) | | 77.1 (49.21) | |
|   Median | 66.2 | 61.7 | | 69.7 | |
|   Min-Max | 6.4-355.4 | 2.8-255 | | 2.8-255 | |
| Number of nodal involvement at treatment initiation –n (%) | | | | | |
|   ≤4 | 39 (40.2) | 74 (51.7) | 0.233 | 38.1 (38.5) | 0.035 |
|   >4 | 58 (59.8) | 69 (48.3) | 0.233 | 60.9 (61.5) | 0.035 |
| Double refractory[b] –n(%) | | | | | |
|   Yes | 66 (68.0) | 97 (67.8) | 0.004 | 67.8 (68.5) | 0.01 |
|   No | 31 (32.0) | 46 (32.2) | 0.004 | 31.2 (31.5) | 0.01 |
| POD24[c] – n(%) | | | | | |
|   Yes | 61 (62.9) | 86 (60.1) | 0.056 | 62.7 (63.3) | 0.009 |
|   No | 36 (37.1) | 57 (39.9) | 0.056 | 36.3 (36.7) | 0.009 |

SMD: standard mean difference; FLIPI: Follicular Lymphoma International Prognostic Index; Auto-HSCT: Autologous haematopoietic stem cell transplantation.

[a] Sample size after weighting (i.e., sum of weights) was 99 for the ReCORD study and effective sample size was 95.

[b] Double refractory status is defined as patients failing to respond or experiencing relapse within 6 months to both a prior anti-CD20 antibody and a prior alkylating agent.

[c] POD24 status is defined as patients failing to respond or experiencing relapse within 24 months to the first-line anti-CD20 mAb containing therapy.

**Table 3:** Efficacy comparison of ELARA and ReCORD before and after weighting

|  | ELARA | Before Weighting ReCORD | After Weighting ReCORD |
|---|---|---|---|
| ***Response rate*** | **N=97** | **N=143** | **N=99**[a] |
| CR, 95% CI | 69.1 (59.8, 78.3) | 39.2 (31.1, 47.2) | 37.3 (26.4, 48.3) |
| ORR, 95% CI | 85.6 (78.7, 92.5) | 68.5 (61.0, 76.1) | 63.6 (52.5, 74.7) |
| Difference in CR, 95% CI |  | 29.9 (17.7, 42.1) | 31.8 (18.1, 45.3) |
| Difference in ORR, 95% CI |  | 17.1 (6.7, 27.4) | 22.0 (9.4, 34.5) |
| **OS** | | | |
| ***Kaplan-Meier Analysis*** | **N=97** | **N=143** | **N=99** |
| Events/Total (%) | 7/97 | 39/143 | 31.3/99 |
| Median, 95% CI (months) | NA | NA | NA |
| 6 months | 100 (100,100) | 88.4 (83.0, 93.7) | 85.6 (77.0, 94.2) |
| 12 months | 96.6 (92.9, 100) | 75.7 (68.3, 83.0) | 71.7 (61.2, 82.2) |
| 18 months | 91.4 (84.6, 98.3) | 71.5 (63.7, 79.3) | 65.8 (54.3, 77.2) |
| 24 months | 87.8 (78.0, 97.6) | 69.7 (61.8, 77.7) | 64.8 (53.3, 76.2) |
| ***Cox proportional hazard model*** | | | |
| HR, 95% CI |  | 0.25 (0.03, 0.46) | 0.20 (0.02, 0.38) |
| **PFS considering new anticancer therapy as event** | | | |
| ***Kaplan-Meier Analysis*** | **N=97** | **N=143** | **N=99** |
| Events/Total (%) | 37/97 | 72/143 | 54.1/99 |
| Median, 95% CI (months) | NA (18.8, NA) | 17.6 (11.2, NA) | 13.1 (8.1, NA) |
| 6 months | 85.3 (78.3, 92.3) | 70.7 (63.1, 78.2) | 66.5 (55.6, 77.3) |
| 12 months | 70.5 (61.4,79.7) | 56.2 (47.9, 64.6) | 51.9 (40.6, 63.3) |
| 18 months | 60.9 (50.4, 71.4) | 48.4 (39.9, 56.9) | 44.6 (33.3, 55.9) |
| 24 months | 54.1 (41.2, 66.9) | 46.7 (38.3, 55.2) | 42.2 (31.0, 53.5) |
| ***Cox proportional hazard model*** | | | |
| HR, 95% CI | | 0.69(0.41,0.97) | 0.60(0.34, 0.86) |

OS and PFS are measured relative to enrollment date/treatment start date. All K-M and Cox regression results are based on survival data within the first 24 months (patients with survival data beyond 24 months were censored at Month 24).

[a] Sample size after weighting (i.e., sum of weights) was 99 for the ReCORD study and effective sample size was 95.

**Table 4.** Efficacy comparison of ELARA and FHRD before and after weighting

|  | ELARA | Before Weighting<br>FHRD | After Weighting[a]<br>FHRD |
|---|---|---|---|
| **Response Rate** | **n=97** | **n=72** | **n=89[b]** |
| CRR (95% CI) | 69.1 (59.8-78.4) | 27.8 (18.1-37.5) | 17.7 (3.8-46.9) |
| ORR (95% CI) | 85.6 (78.4-91.8) | 62.5 (51.4-73.6) | 58.1 (21.3-88.2) |
| Difference in CRR (95% CI) |  | 41.3 (27.1-55.1) | 51.4 (21.2-68.8) |
| Difference in ORR (95% CI) |  | 23.1 (9.9-35.9) | 27.4 (-3-65) |
| **OS** |  |  |  |
| *Kaplan-Meier Analysis* | **n=97** | **n=98** | **n=88[c]** |
| Events/Total (%) | 7/97 (7.2%) | 24/98 (24.5%) | 12.8/88 (14.5%) |
| Median, 95% CI (months) | NR | NR | NR |
| 6 months | 100.0 (100.0-100.0) | 86.7 (79.3-93.4) | 95.7 (88.7-99.2) |
| 12 months | 96.6 (92.3-100.0) | 77.0 (67.5-86) | 84.5 (64.9-95.9) |
| 18 months | 91.4 (84.1-97.6) | 71.8 (61.4-81.6) | 82.7 (62.6-94.6) |
| 24 months | 87.8 (77.3-96.2) | 67.7 (56.5-78.5) | 79.1 (58.8-92.5) |
| *Cox proportional hazard model* |  |  |  |
| HR, 95% CI, Study E2202 vs Flatiron (Ref) |  | 0.24 (0.08-0.51) | 0.41 (0.11-1.47) |
| **PFS** |  |  |  |
| *Kaplan-Meier Analysis* | **n=97** | **n=98** | **n=88[c]** |
| Events/Total (%) | 34/97 (35.1%) | 52/98 (53.1%) | 48.3/88 (54.8%) |
| Median, 95% CI (months) | NR | 9.9 (6.8-19.3) | 9.9 (8-19.3) |
| 6 months | 87.2 (80.2-93.6) | 63.7 (53.6-73.7) | 77 (60.5-88.5) |
| 12 months | 73.2 (64.1-82.1) | 45.2 (34.1-56.5) | 41.8 (20-67.2) |
| 18 months | 63.2 (52.4-73.5) | 39.0 (27.4-50.6) | 29.8 (11.6-56.2) |
| 24 months | 56.1 (41.8-68.9) | 32.6 (20.9-44.4) | 26.2 (8.1-52.0) |
| *Cox proportional hazard model* |  |  |  |

| | | |
|---|---|---|
| HR, 95% CI | 0.45 (0.29-0.69) | 0.45 (0.26-0.88) |

*10,000 bootstrap samples were randomly drawn to calculate the percentile-based 95% CI

*All K-M and cox regression results are based on survival data within the first 24 months (patients with survival data beyond 24 months were censored at Month 24).

[a] Only patients with at least one evaluation for response or a documented death during treatment were considered for this analysis.

[b] Sample size after weighting (i.e., sum of weights) was 89 for the FHRD study and effective sample size was 18.

[c] Sample size after weighting (i.e., sum of weights) was 88 for the FHRD study and effective sample size was 29.

**Figure 1.** Elara trial design

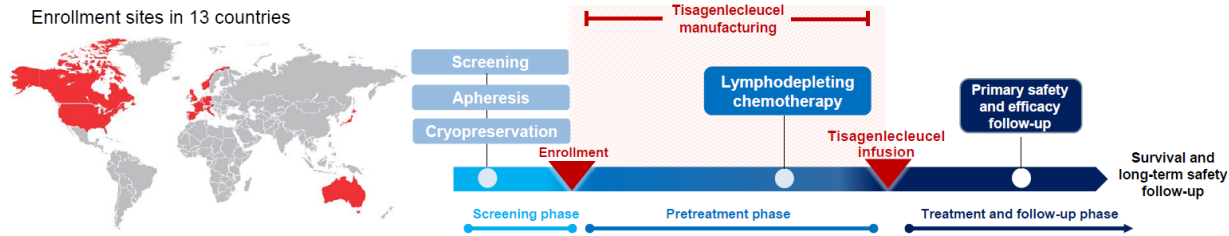

**Figure 2.** Propensity score distribution in: a) the original ReCORD sample (before weighting); and b) the weighted sample

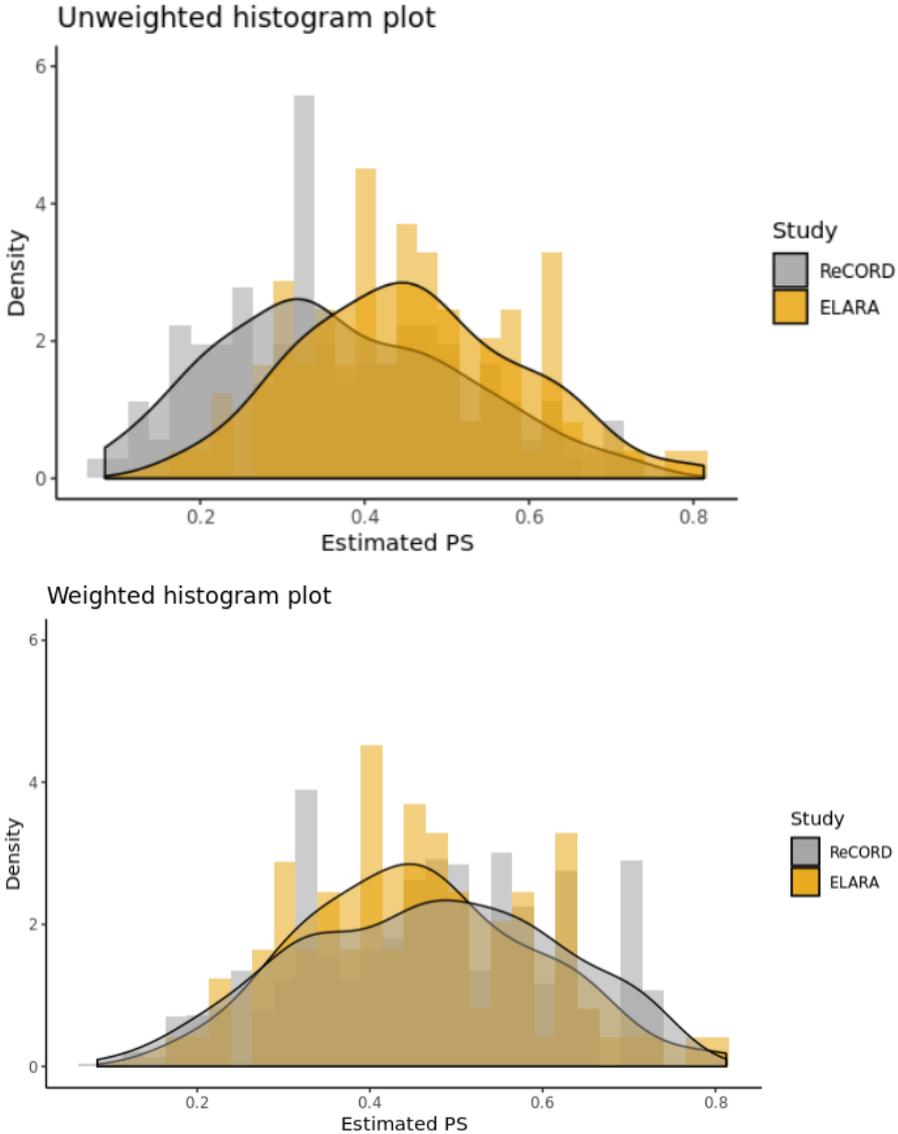

**Figure 3.** Kaplan-Meier plots of: a) OS; and b) PFS for ELARA and ReCORD after weighting adjustment

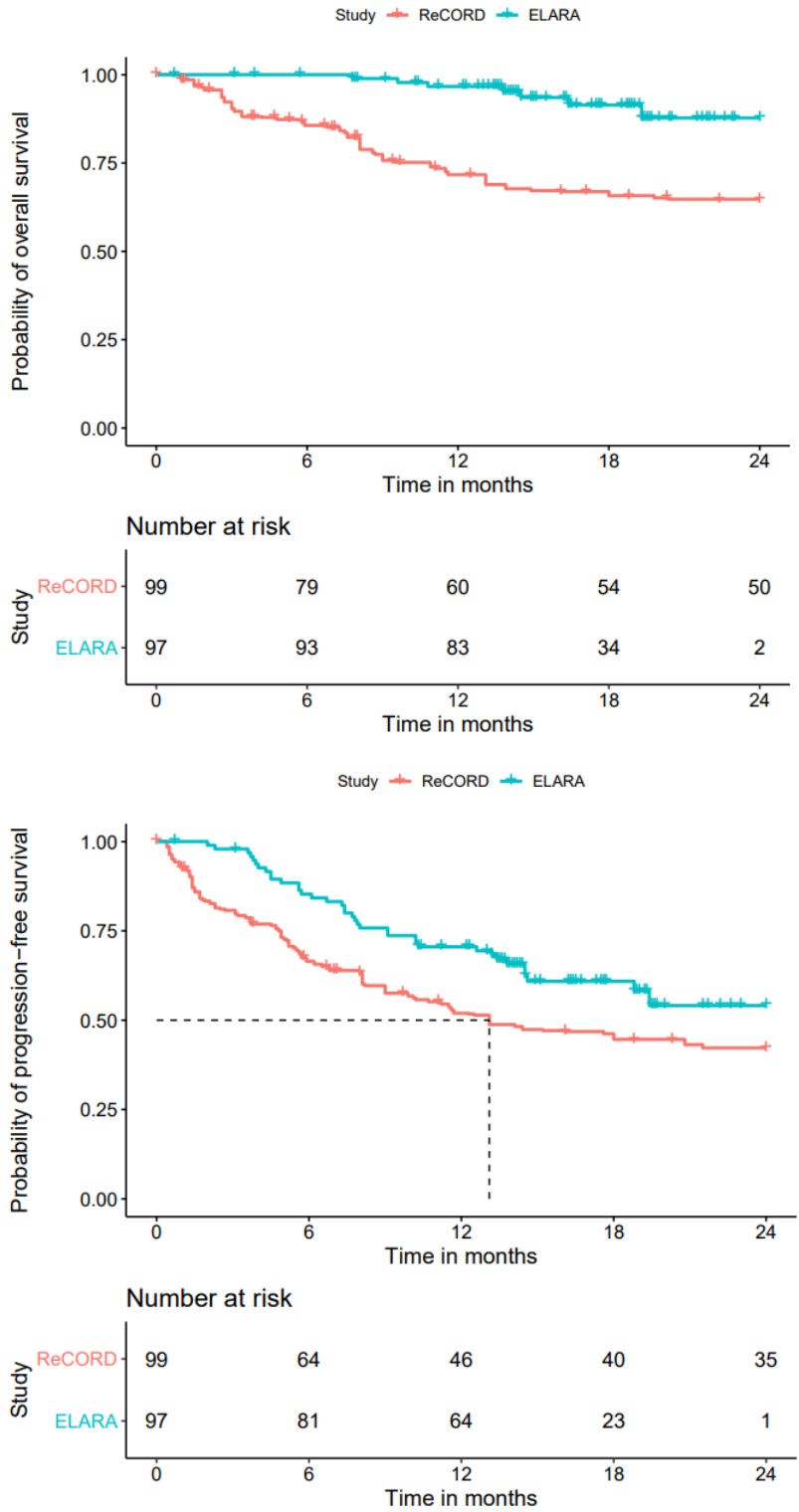

**Figure 4.** Propensity score distribution in: a) the original FHRD sample (before weighting); and b) the weighted sample

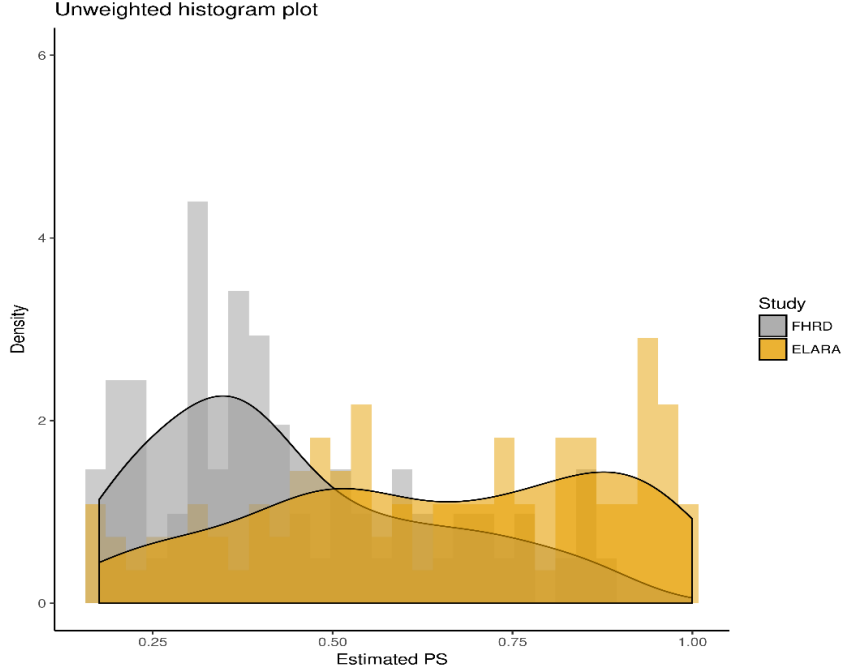

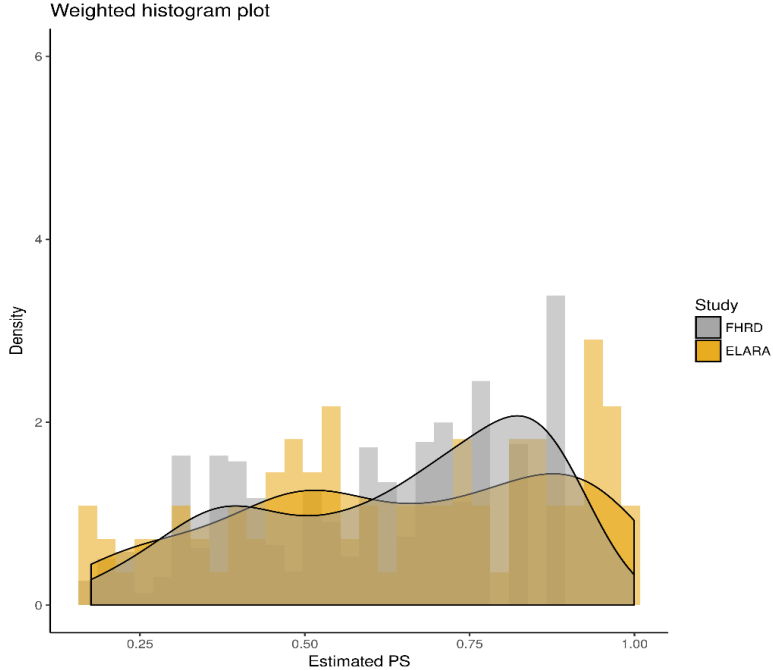